\documentclass[twocolumn,conference]{IEEEtran}
\usepackage[T1]{fontenc}
\usepackage[latin9]{inputenc}
\usepackage{amsmath}
\usepackage{amsthm}
\usepackage{amssymb}
\usepackage{graphicx}
\usepackage[unicode=true,
 bookmarks=true,bookmarksnumbered=true,bookmarksopen=true,bookmarksopenlevel=1,
 breaklinks=false,pdfborder={0 0 0},pdfborderstyle={},backref=false,colorlinks=false]
 {hyperref}
\hypersetup{pdftitle={Your Title},
 pdfauthor={Your Name},
 pdfpagelayout=OneColumn, pdfnewwindow=true, pdfstartview=XYZ, plainpages=false}

\makeatletter

\providecommand{\tabularnewline}{\\}

\theoremstyle{plain}
\newtheorem{thm}{\protect\theoremname}

\usepackage[caption=false,font=footnotesize]{subfig}

\makeatother

\providecommand{\theoremname}{Theorem}

\begin{document}

\title{Time-Domain Multi-Beam Selection and Its Performance Improvement
for mmWave Systems}

\author{\IEEEauthorblockN{Hsiao-Lan~Chiang, Wolfgang~Rave, and Gerhard~Fettweis}\IEEEauthorblockA{Vodafone Chair for Mobile Communications, Technische Universität Dresden,
Germany\\
Email: \{hsiao-lan.chiang, rave, gerhard.fettweis\}@ifn.et.tu-dresden.de}}
\maketitle
\begin{abstract}
Multi-beam selection is one of the crucial technologies in hybrid
beamforming systems for frequency-selective fading channels. Addressing
the problem in the \textit{frequency} domain facilitates the procedure
of acquiring observations for analog beam selection. However, it is
difficult to improve the quality of the contaminated observations
at low SNR. To this end, this paper uses an idea that the significant
observations are sparse in the \textit{time} domain to further enhance
the quality of signals as well as the beam selection performance.
In OFDM systems, by exploiting properties of channel impulse responses
and circular convolutions in the time domain, we can reduce the size
of a circulant matrix in deconvolution to generate periodic true values
of coupling coefficients plus random noise signals. An arithmetic
mean of these signals yields refined observations with minor noise
effects and provides more accurate sparse multipath delay information.
As a result, only the refined observations associated with the estimated
multipath delay indices have to be taken into account for the analog
beam selection problem.
\end{abstract}

\section{Introduction}

With the rapid increase of data rates in wireless communications,
bandwidth shortage is getting more critical. Accordingly, there is
a growing interest in using millimeter wave (mmWave) for future wireless
communications taking advantage of the enormous amount of available
spectrum \cite{Rappaport2014}. In mmWave systems, a combination of
analog beamforming (operating in passband) \cite{Liberti1999,Hajimiri2005}
and digital beamforming (operating in baseband) \cite{Veen1988} is
one of the low-cost solutions to higher data rate transmission, and
this combination is commonly called hybrid beamforming \cite{Zhang2005}-\nocite{Ayach2014}\nocite{Yu2015}\cite{Han2015}.
To implement hybrid beamforming at a transmitter and a receiver simultaneously
is certainly intractable. Therefore, our previous works in \cite{Chiang2017_JSTSP,Chiang2018_ICC_FD}
focus on finding the key parameters of the hybrid beamforming gain
to alleviate the problem, and eventually all that matters about the
hybrid beamforming performance is the analog beam selection. 

The problem of analog beam selection for frequency-selective fading
channels can be stated as a sum-power (or energy) maximization across
all subcarriers \cite{Chiang2018_ICC_FD,Alkhateeb2016b}. From Parseval's
theorem, we know that it is equivalent to calculating the energy of
the observations for the analog beam selection in the delay (or time)
domain. Particularly, the observations in the delay domain can be
interpreted as coupling coefficients of a matrix-valued channel impulse
response (CIR) and all possible analog beam pairs plus noise. Considering
an OFDM system, it is easier to obtain the observations in the frequency
domain. However, these signals seriously suffer from the noise in
the low SNR regime, and it needs more effort to refine them in the
frequency domain than in the delay domain because the significant
observations are not sparse in the frequency domain. To this end,
this paper presents a low-complexity beam selection method and its
performance improvement in the delay domain.

In OFDM systems, the delay-domain convolution operation can be constructed
as a matrix multiplication, where one of the inputs (that is, the
training sequence) is converted into a circulant matrix. Then, left
multiplying the received signal vector by the inverse of the circulant
matrix leads to the observations for the analog beam selection. 
In the system, the length $L_{C}$ of a cyclic prefix (CP) is much
less than one OFDM symbol duration with $L$ samples but is enough
to cover the maximum delay spread \cite{Tse2005}, which means that
at most $L_{C}$ observations in one OFDM symbol can be used for the
beam selection. Unfortunately, the $L_{C}$ observations are unreliable
in the low SNR regime.

In order to improve the quality of the observations for the beam
selection, we generate the training sequence of length $L_{C}$ with
a certain period $M=\left\lfloor \tfrac{L}{L_{C}}\right\rfloor $
within one OFDM symbol duration at the transmitter. After deconvolution
by a small-size circulant matrix, we have $M$ periodic signals of
length $L_{C}$ plus random noise signals. An arithmetic mean of these
signals yields the refined observations, where the effective noise
variance is reduced by a factor of $M$. According to one of the transmission
numerologies in 3GPP 5G New Radio (NR) \cite{3GPP38211}, $M\approx14$
so that the mean absolute error (MAE) between the energy estimate
and its true value can be significantly reduced. In addition, if the
refined observations are reliable enough to find the delay indices,
eventually only a few number of signals corresponding to the estimated
delay indices are the significant observations for the analog beam
selection.

The following notations are used throughout this paper. $a$ is
a scalar, $\mathbf{a}$ is a column vector, and $\mathbf{A}$ is a
matrix. \textbf{$\mathbf{a}_{n}$} denotes the $n^{\text{th}}$ column
vector of $\mathbf{A}$; $a_{i,j}$ denotes the $(i,j)^{\text{th}}$
entry of $\mathbf{A}$. $\mathbf{A}^{T}$ and $\mathbf{A}^{H}$ denote
the transpose and Hermitian transpose of $\mathbf{A}$ respectively.
$\left[\mathbf{A}\right]_{n,:}$ denotes the $n^{\text{th}}$ row
vector of $\mathbf{A}$. $\mathbf{I}_{N}$ and $\mathbf{0}_{N\times M}$
denote respectively the $N\times N$ identity and $N\times M$ zero
matrices. $a[l]\circledast_{L}b[l]$ denotes the circular convolution
of sequences $a[l]$ and $b[l]$ of length $L$.

\begin{figure}[t]
\begin{centering}
\includegraphics[scale=0.55]{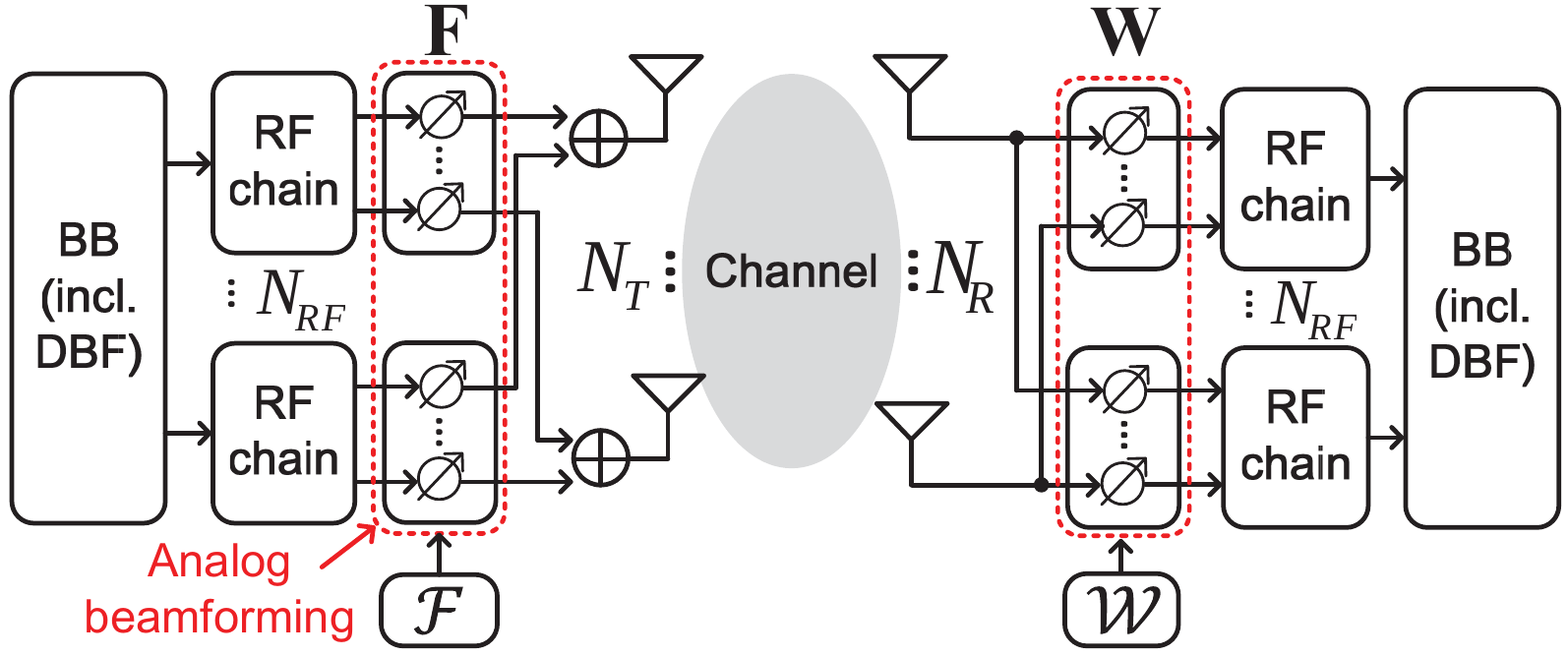}
\par\end{centering}
\centering{}\caption{Both a transmitter and a receiver have $N_{RF}$ analog beamforming
vectors and a baseband (BB) signal processing block including digital
beamforming (DBF). This paper focuses on an analog beam selection
problem, which dominates the complexity and performance of hybrid
beamforming systems \cite{Chiang2018_ICC_FD}. \label{fig:System-diagram.} }
\end{figure}

\section{System Model}

A system having a transmitter with an $N_{T}$-element uniform linear
antenna array (ULA) communicates $N_{RF}$ data streams to a receiver
with an $N_{R}$-element ULA as shown in Fig. \ref{fig:System-diagram.}.
The $N_{RF}$ analog beamforming vectors at the transmitter in matrix
$\mathbf{F}=[\mathbf{f}_{1},\cdots,\mathbf{f}_{N_{RF}}]$ are selected
from a predefined codebook $\mathcal{F}=\{\tilde{\mathbf{f}}_{n_{f}}\in\mathbb{C}^{N_{T}\times1},n_{f}=1,\cdots,N_{F}\}$
with the $n_{f}^{\text{th}}$ member represented as \cite{Liberti1999}
\begin{multline}
\tilde{\mathbf{f}}_{n_{f}}=\frac{1}{\sqrt{N_{T}}}\left[1,e^{j\frac{2\pi}{\lambda_{0}}\text{sin}(\phi_{T,n_{f}})\Delta_{d}},\cdots,\right.\\
\left.e^{j\frac{2\pi}{\lambda_{0}}\text{sin}(\phi_{T,n_{f}})\cdot(N_{T}-1)\Delta_{d}}\right]^{T},\label{eq: f}
\end{multline}
where $\phi_{T,n_{f}}$ stands for the $n_{f}^{\text{th}}$ candidate
of the steering angles at the transmitter, $\Delta_{d}=\tfrac{\lambda_{0}}{2}$
is the distance between two neighboring antennas, and $\lambda_{0}$
is the wavelength at the carrier frequency. At the receiver, the $N_{RF}$
analog beamforming vectors in matrix $\mathbf{W}=[\mathbf{w}_{1},\cdots,\mathbf{w}_{N_{RF}}]$
are selected from the other codebook defined as $\mathcal{W}=\{\tilde{\mathbf{w}}_{n_{w}}\in\mathbb{C}^{N_{R}\times1},n_{w}=1,\cdots,N_{W}\}$,
where the members can be generated by the same rule as (\ref{eq: f}).
The analog beamforming matrices $\mathbf{F}$ and $\mathbf{W}$ are
assumed to be constant within one OFDM symbol duration owing to hardware
constraints.

Via a coupling of two analog beamforming matrices and a multipath
matrix-valued CIR $\mathbf{H}[l]\in\mathbb{C}^{N_{R}\times N_{T}}$,
where $l=0,\cdots,L-1$ denotes the sample in one OFDM symbol, the
$l^{\text{th}}$ sampled received signal vector $\mathbf{r}[l]\in\mathbb{C}^{N_{RF}\times1}$
can be written as
\begin{equation}
\begin{alignedat}{1}\mathbf{r}[l] & =\sqrt{\rho}\cdot\mathbf{W}^{H}\mathbf{H}[l]\circledast_{L}\mathbf{F}\mathbf{s}[l]+\mathbf{W}^{H}\mathbf{n}[l],\end{alignedat}
\label{eq: r}
\end{equation}
where $\rho$ stands for the average received power containing the
transmit power, transmit antenna gain, receive antenna gain, and path
loss, $\mathbf{s}[l]\in\mathbb{C}^{N_{RF}\times1}$ is the transmitted
signal vector, and $\mathbf{n}[l]\in\mathbb{C}^{N_{R}\times1}$ is
an $N_{R}$-dimensional independent and identically distributed (i.i.d.)
complex Gaussian random vector, $\mathbf{n}[l]\sim\mathcal{CN}(\boldsymbol{0}_{N_{R}\times1},\sigma_{n}^{2}\mathbf{I}_{N_{R}})$. 

mmWave channel models have been widely studied recently \cite{3GPP38900,Rappaport2015}.
Based on the references, a simplified mmWave CIR matrix $\mathbf{H}[l]$
can be expressed as the sum of $P$ outer products of the array response
vectors associated with the normalized-quantized delay $l_{p}=\left\lfloor \tau_{p}F_{S}\right\rfloor \in\mathbb{N}_{0}$
(the set of natural numbers contains zero), where $\tau_{p}\in\mathbb{R}_{\geq0}$
(the set of positive real numbers contains zero) is the delay for
path $p$ and $F_{S}$ is the sampling rate,
\begin{equation}
\begin{alignedat}{1}\mathbf{H}[l] & =\sum_{p=1}^{P}\underset{c_{p}[l]}{\underbrace{\alpha_{p}\delta[l-l_{p}]}}\cdot\mathbf{a}_{A}(\phi_{A,p})\mathbf{a}_{D}(\phi_{D,p})^{H}\\
 & =\sum_{p=1}^{P}c_{p}[l]\mathbf{a}_{A}(\phi_{A,p})\mathbf{a}_{D}^{H}(\phi_{D,p})
\end{alignedat}
\label{eq: H_l}
\end{equation}
where $\alpha_{p}\in\mathbb{R}_{>0}$ is the attenuation coefficient
for path $p$ and $\sum_{p=1}^{P}|\alpha_{p}|^{2}=1$. Note that the
path loss values influenced by an environment and geometry are mentioned
in the average received power $\rho$ in (\ref{eq: r}). $c_{p}[l]$
characterizes the CIR for path $p$ at sample $l$ and we assume that
$c_{p}[l]=0$ when $l\geq L_{C}$, where $L_{C}$ is the CP length.
The departure array response vector $\mathbf{a}_{D}(\phi_{D,p})$
is a function of angle of departure (AoD), $\phi_{D,p}\sim\mathcal{U}(-\tfrac{\pi}{2},\tfrac{\pi}{2})$,
for path $p$,
\begin{multline}
\mathbf{a}_{D}(\phi_{D,p})=\frac{1}{\sqrt{N_{T}}}\left[1,e^{j\tfrac{2\pi}{\lambda_{0}}\text{sin}(\phi_{D,p})\Delta_{d}},\cdots,\right.\\
\left.e^{j\tfrac{2\pi}{\lambda_{0}}\text{sin}(\phi_{D,p})(N_{T}-1)\Delta_{d}}\right]^{T},\label{eq: a_d}
\end{multline}
and the arrival array response vector $\mathbf{a}_{A}(\phi_{A,p})$,
where $\phi_{A,p}\sim\mathcal{U}(-\tfrac{\pi}{2},\tfrac{\pi}{2})$,
has a similar form as (\ref{eq: a_d}).

\section{Time-Domain Analog Beam Selection}

\subsection{Observations for analog beam selection}

In order to acquire the observations for the analog beam selection,
we simply assume that all the beam pairs selected from $\mathcal{F}$
and $\mathcal{W}$ are trained by a known training sequence. Hypothetically
there is no data transmission and reception before the transmitter
and receiver select the preferable analog beam pairs. Hence, one can
use a training sequence of length $L$ in one OFDM symbol, $\{s[0],\cdots,s[L-1]\}$,
to train one beam pair. The $l^{\text{th}}$ sampled scalar of the
received signals by using the beam pair $(\tilde{\mathbf{f}}_{n_{f}},\tilde{\mathbf{w}}_{n_{w}})$
can therefore be expressed as
\begin{equation}
\begin{alignedat}{1}r_{n_{w},n_{f}}[l] & =\sqrt{\rho}\cdot\tilde{\mathbf{w}}_{n_{w}}^{H}\mathbf{H}[l]\circledast_{L}\tilde{\mathbf{f}}_{n_{f}}s[l]+\underset{z_{n_{w},n_{f}}[l]}{\underbrace{\tilde{\mathbf{w}}_{n_{w}}^{H}\mathbf{n}[l]}}\\
 & =\sqrt{\rho}\cdot\tilde{\mathbf{w}}_{n_{w}}^{H}\mathbf{H}[l]\circledast_{L}\tilde{\mathbf{f}}_{n_{f}}s[l]+z_{n_{w},n_{f}}[l],
\end{alignedat}
\label{eq: r=00005Bl=00005D}
\end{equation}
where $n_{f}=1,\cdots,N_{F}$, $n_{w}=1,\cdots,N_{W}$, the combined
noise $z_{n_{w},n_{f}}[l]\sim\mathcal{CN}(0,\sigma_{n}^{2})$ still
has a Gaussian distribution with mean zero and variance $\sigma_{n}^{2}$
due to the equal-magnitude elements of $\tilde{\mathbf{w}}_{n_{w}}$.

To implement deconvolution of the received signal and get the observations
for the beam selection, we intend to decouple the angle- and delay-domain
components in $r_{n_{w},n_{f}}[l]$ by replacing the channel matrix
$\mathbf{H}[l]$ with (\ref{eq: H_l}). Consequently, $r_{n_{w},n_{f}}[l]$
can be further written as follows:

{\small{}\vspace*{-0.2cm}
\begin{equation}
\begin{alignedat}{1} & r_{n_{w},n_{f}}[l]\\
 & =\sqrt{\rho}\cdot\sum_{p=1}^{P}\underset{\triangleq\eta_{p,n_{w},n_{f}}}{\underbrace{\tilde{\mathbf{w}}_{n_{w}}^{H}\mathbf{a}_{A}(\phi_{A,p})\mathbf{a}_{D}^{H}(\phi_{D,p})\tilde{\mathbf{f}}_{n_{f}}}}\left(c_{p}[l]\circledast_{L}s[l]\right)+z_{n_{w},n_{f}}[l]\\
 & =\sqrt{\rho}\cdot\sum_{p=1}^{P}\eta_{p,n_{w},n_{f}}\cdot\left(c_{p}[l]\circledast_{L}s[l]\right)+z_{n_{w},n_{f}}[l],
\end{alignedat}
\label{eq: r=00005Bl=00005D-1}
\end{equation}
}where $|\eta_{p,n_{w},n_{f}}|=|\tilde{\mathbf{w}}_{n_{w}}^{H}\mathbf{a}_{A}(\phi_{A,p})|\cdot|\mathbf{a}_{D}^{H}(\phi_{D,p})\tilde{\mathbf{f}}_{n_{f}}|$
is the multiplication of beamforming gains at the transmitter and
receiver.

Then, we collect $L$ samples in a vector and express the circular
convolution as a multiplication by a circulant matrix $\mathbf{S}$
\cite{Kay1997}
\begin{equation}
\begin{alignedat}{1}\mathbf{r}_{n_{w},n_{f}} & =\left[r_{n_{w},n_{f}}[0],\cdots,r_{n_{w},n_{f}}[L-1]\right]^{T}\\
 & =\sqrt{\rho}\cdot\mathbf{S}\sum_{p=1}^{P}\eta_{p,n_{w},n_{f}}\mathbf{c}_{p}+\mathbf{z}_{n_{w},n_{f}},
\end{alignedat}
\label{eq: vector_r}
\end{equation}
where\begin{subequations}
\begin{align}
\mathbf{S} & =\left[\begin{array}{ccc}
s[0] & \cdots & s[1]\\
\vdots & \ddots & \vdots\\
s[L-1] & \cdots & s[0]
\end{array}\right]\in\mathbb{C}^{L\times L},\\
\mathbf{c}_{p} & =\left[c_{p}[0],\cdots,c_{p}[L-1]\right]^{T}\in\mathbb{C}^{L\times1},\\
\mathbf{z}_{n_{w},n_{f}} & =\left[z_{n_{w},n_{f}}[0],\cdots,z_{n_{w},n_{f}}[L-1]\right]^{T}\in\mathbb{C}^{L\times1}.
\end{align}
\end{subequations}The $L$ observations can therefore be obtained
by pre-multiplying $\mathbf{r}_{n_{w},n_{f}}$ by $\mathbf{S}^{-1}$,
where $\det(\mathbf{S})\neq0$, given by
\begin{equation}
\begin{alignedat}{1}y_{n_{w},n_{f}}[l] & =\left[\mathbf{S}^{-1}\right]_{l,:}\mathbf{r}_{n_{w},n_{f}}\\
 & =\sqrt{\rho}\cdot\sum_{p=1}^{P}\eta_{p,n_{w},n_{f}}c_{p}[l]+\underset{\xi_{n_{w},n_{f}}[l]}{\underbrace{\left[\mathbf{S}^{-1}\right]_{l,:}\mathbf{z}_{n_{w},n_{f}}}}\\
 & =\sqrt{\rho}\cdot\tilde{\mathbf{w}}_{n_{w}}^{H}\mathbf{H}[l]\tilde{\mathbf{f}}_{n_{f}}+\xi_{n_{w},n_{f}}[l],
\end{alignedat}
\label{eq: y}
\end{equation}
where $l=0,\cdots,L-1$. One can design the training sequence so that
$\xi_{n_{w},n_{f}}[l]$ has a complex Gaussian distribution with mean
zero and a variance of $\sigma_{\xi}^{2}$.

\subsection{Problem statement\label{subsec:Problem-statement}}

The observations $\{y_{n_{w},n_{f}}[l]\:\forall n_{w},n_{f},l\}$
can be interpreted as coupling coefficients of the channel and the
trained beam pairs. If the coupling coefficients are acquired in the
frequency domain, our previous work in \cite{Chiang2018_ICC_FD} introduces
how to use them to select the analog beam pairs. Simply speaking,
the problem of frequency-domain analog beam selection can be formulated
as finding the beam pairs that maximize the sum of the power of the
observations across all subcarriers. From Parseval's theorem, we know
that the objective function is equivalent to the sum of the power
across all samples in the delay domain. As a result, the delay-domain
analog beam selection can be expressed as the following maximization
problem:

\begin{equation}
\begin{gathered}(\hat{\mathbf{f}}_{n_{rf}},\hat{\mathbf{w}}_{n_{rf}})={\displaystyle \underset{\scriptsize\begin{array}{c}
\tilde{\mathbf{f}}_{n_{f}}\in\mathcal{F}\backslash\mathcal{F}',\tilde{\mathbf{w}}_{n_{w}}\in\mathcal{W}\backslash\mathcal{W}'\end{array}}{\arg\,\max}}g_{n_{w},n_{f}},\end{gathered}
\label{eq: ABF}
\end{equation}
where $n_{rf}=1,\cdots,N_{RF}$, $g_{n_{w},n_{f}}$ is the energy
of the observations
\begin{equation}
\begin{alignedat}{1}g_{n_{w},n_{f}} & =\sum_{l=0}^{L-1}\left|y_{n_{w},n_{f}}[l]\right|^{2}\end{alignedat}
,\label{eq: g_ori}
\end{equation}
$\mathcal{F}'=\{\hat{\mathbf{f}}_{n},n=1,\cdots,n_{rf}-1\}$ and $\mathcal{W}'=\{\hat{\mathbf{w}}_{n},n=1,\cdots,n_{rf}-1\}$
are the sets including the selected analog beamforming vectors from
iteration $1$ to $n_{rf}-1$. The energy estimate $g_{n_{w},n_{f}}$
is also the objective function used in frequency-domain analog beam
selection problem \cite{Chiang2018_ICC_FD}. However, in the frequency
domain, we do not have the information that $y_{n_{w},n_{f}}[l]$,
$l=L_{C},\cdots,L-1$, only contain noise signals.

In the beam selection problem stated in (\ref{eq: ABF}), the sum
of the power of $L$ noise-free observations, i.e.,
\begin{equation}
y_{n_{w},n_{f}}^{N\!F}[l]\triangleq\sqrt{\rho}\,\tilde{\mathbf{w}}_{n_{w}}^{H}\mathbf{H}[l]\tilde{\mathbf{f}}_{n_{f}},\label{eq: y_NF}
\end{equation}
where $l=0,\cdots,L-1$, would lead to the optimal solution. Let us
write down the corresponding objective function
\begin{equation}
\begin{alignedat}{1}g_{n_{w},n_{f}}^{N\!F} & \triangleq\sum_{l=0}^{L-1}\left|y_{n_{w},n_{f}}^{N\!F}[l]\right|^{2}\\
 & =\sum_{p=1}^{P}\left|\sqrt{\rho}\,\tilde{\mathbf{w}}_{n_{w}}^{H}\mathbf{H}[l_{p}]\tilde{\mathbf{f}}_{n_{f}}\right|^{2},
\end{alignedat}
\label{eq: g_NF}
\end{equation}
where the second equality follows from that $\mathbf{H}[l]=0$ when
$l\notin\{l_{p},p=1,\cdots,P\}$ and $l_{p}$, $p=1,\cdots,P$, are
different to each other. Compared with (\ref{eq: g_ori}), it is clear
that in (\ref{eq: g_NF}) only $P$ (rather than $L$) observations
associated with the $P$ delay indices have to be taken into account.
Therefore, our goal is to reduce the error between the energy estimate
$g_{n_{w},n_{f}}$ and its true value $g_{n_{w},n_{f}}^{N\!F}$ without
an additional computational overhead. 

\section{Performance Improvement of Analog Beam Selection in Time Domain\label{sec:Performance-Improvement-of}}

\subsection{Performance metric}

From the discussion in the previous subsection, we know that there
is a higher probability to find the optimal solution when the error
between $g_{n_{w},n_{f}}$ and $g_{n_{w},n_{f}}^{N\!F}$ approximates
to zero. Therefore, we use the MAE between $g_{n_{w},n_{f}}$ and
$g_{n_{w},n_{f}}^{N\!F}$ as a performance metric to quantify the
performance of beam selection, which is stated in \textbf{Theorem
\ref{thm: MAE}}. We consider the MAE rather than the mean squared
error (MSE) due to that fact that $g_{n_{w},n_{f}}$ and $g_{n_{w},n_{f}}^{N\!F}$
are energy signals; it is redundant to calculate the \textit{squared}
error between these two values.
\begin{thm}
\label{thm: MAE}Given matrix-valued CIRs $\underline{\mathbf{H}}[l]$,
$l=0,\cdots,L-1$, one has the energy estimates 
\begin{equation}
\underline{g}_{n_{w},n_{f}}=\sum_{l=0}^{L-1}\left|\sqrt{\rho}\,\tilde{\mathbf{w}}_{n_{w}}^{H}\underline{\mathbf{H}}[l]\tilde{\mathbf{f}}_{n_{f}}+\xi_{n_{w},n_{f}}[l]\right|^{2}\,\forall n_{w},n_{f}
\end{equation}
and the corresponding true values 
\begin{equation}
\underline{g}_{n_{w},n_{f}}^{N\!F}=\sum_{p=1}^{P}\left|\sqrt{\rho}\,\tilde{\mathbf{w}}_{n_{w}}^{H}\underline{\mathbf{H}}[l_{p}]\tilde{\mathbf{f}}_{n_{f}}\right|^{2}\,\forall n_{w},n_{f}.
\end{equation}
Then the MAE between $\underline{g}_{n_{w},n_{f}}$ and $\underline{g}_{n_{w},n_{f}}^{N\!F}$
is upper bounded by
\begin{equation}
\begin{alignedat}{1}\text{MAE}(\underline{g}_{n_{w},n_{f}}) & \triangleq\text{E}\left[\left|\underline{g}_{n_{w},n_{f}}-\underline{g}_{n_{w},n_{f}}^{N\!F}\right|\right]\\
 & \leq\text{E}\left[\left|\varepsilon_{n_{w},n_{f}}\right|\right]+\text{E}\left[\nu\right],
\end{alignedat}
\label{eq:  MAE}
\end{equation}
where 
\begin{equation}
\varepsilon_{n_{w},n_{f}}\sim\mathcal{N}\left(0,2\sigma_{\xi}^{2}\underline{g}_{n_{w},n_{f}}^{N\!F}\right)\label{eq: epsilon}
\end{equation}
 and 
\begin{equation}
\nu\sim\Gamma(L,\sigma_{\xi}^{2}).\label{eq: nu}
\end{equation}
\end{thm}
\begin{IEEEproof}
See Appendix A.
\end{IEEEproof}
$ $

\subsection{Refine observations by averaging random noise signals}

In OFDM systems, the CP length ($L_{C}$) is designed to cover the
maximum or root-mean-square (RMS) delay spread, which means that
the number of useful observations in one OFDM symbol is less than
or equal to $L_{C}$. To improve the quality of the observations,
we use a property of circular convolutions introduced as follows.
First, simply modifying the transmitted training sequence of length
$L$ as $M=\tfrac{L}{L_{C}}$ (assume $\tfrac{L}{L_{C}}\in\mathbb{N}^{+}$)
repeated sequence blocks, where the length of each block is $L_{C}$.
Such periodic training sequence blocks make the circular convolution
in (\ref{eq: r=00005Bl=00005D-1}) become
\begin{equation}
\begin{alignedat}{1}c_{p}[l]\circledast_{L}s[l] & =\sum_{n=0}^{L-1}c_{p}[n]s[l-n]\\
 & =\sum_{n=0}^{L_{C}-1}c_{p}[n]s[l-n]\\
 & =c_{p}[l]\circledast_{L_{C}}s[l],
\end{alignedat}
\end{equation}
where the second equality follows from that $c_{p}[l]=0$ when $l\geq L_{C}$,
and $\circledast_{L_{C}}$ denotes a circular convolution over the
cyclic group of integers modulo $L_{C}$. Then, following from (\ref{eq: y}),
we can use a circulant matrix of small size $L_{C}\times L_{C}$ (generated
by one training sequence block) to sequentially implement the deconvolution
of $M$ received periodic blocks. An arithmetic mean of the $M$ outputs
of the deconvolution leads to a result suffering from less noise effect
\begin{equation}
\begin{alignedat}{1}y_{n_{w},n_{f}}'[l_{c}] & =y_{n_{w},n_{f}}^{N\!F}[l_{c}]+\underset{\xi_{n_{w},n_{f}}'[l_{c}]}{\underbrace{\frac{1}{M}\sum_{m=1}^{M}\xi_{n_{w},n_{f}}[(m-1)L_{C}+l_{c}]}}\\
 & =y_{n_{w},n_{f}}^{N\!F}[l_{c}]+\xi_{n_{w},n_{f}}'[l_{c}],
\end{alignedat}
\label{eq: y_avg}
\end{equation}
where $l_{c}=0,\cdots,L_{C}-1$, and the variance of $\xi_{n_{w},n_{f}}'[l_{c}]\sim\mathcal{CN}\left(0,\frac{\sigma_{\xi}^{2}}{M}\right)$
is effectively reduced by a factor of $M$. By using these averaged
(or refined) observations, the energy estimate in (\ref{eq: g_ori})
becomes
\begin{equation}
g_{n_{w},n_{f}}'=\sum_{l_{c}=0}^{L_{C}-1}\left|y_{n_{w},n_{f}}'[l_{c}]\right|^{2}.\label{eq: g_avg}
\end{equation}

Based on the\textbf{ }derivation of \textbf{Theorem \ref{thm: MAE}},
the MAE between the estimate $g_{n_{w},n_{f}}'$ and its true value
$g_{n_{w},n_{f}}^{N\!F}$ conditioned on the same channel realizations,
$\underline{\mathbf{H}}[l]$, $l=0,\cdots,L-1$, is given by
\begin{equation}
\begin{alignedat}{1}\text{MAE}\left(\underline{g}_{n_{w},n_{f}}'\right) & \triangleq\text{E}\left[\left|\underline{g}_{n_{w},n_{f}}'-\underline{g}_{n_{w},n_{f}}^{N\!F}\right|\right]\\
 & \leq\text{E}\left[\left|\varepsilon_{n_{w},n_{f}}'\right|\right]+\text{E}\left[\nu'\right],
\end{alignedat}
\label{eq: MAE_avg}
\end{equation}
where $\varepsilon_{n_{w},n_{f}}'$ has a Gaussian distribution
\begin{equation}
\varepsilon_{n_{w},n_{f}}'\sim\mathcal{N}\left(0,2\left(\frac{\sigma_{\xi}^{2}}{M}\right)\underline{g}_{n_{w},n_{f}}^{N\!F}\right),\label{eq:epsilon_avg}
\end{equation}
and $\nu'$ follows a gamma distribution
\begin{equation}
\nu'\sim\Gamma\left(\frac{L}{M},\frac{\sigma_{\xi}^{2}}{M}\right).\label{eq:nu_avg}
\end{equation}
Compared with (\ref{eq: epsilon}), (\ref{eq: nu}), the noise effect
caused by $\varepsilon_{n_{w},n_{f}}'$ and $\nu'$ can be effectively
reduced when $M$ is large. For example, one of the use cases in 3GPP
5G NR \cite{3GPP38211} shows that the CP ratio $\tfrac{1}{M}\approx\tfrac{1}{14}$.

\subsection{Further refine observations by using knowledge of multipath delay\label{subsec:Further-refine-observations}}

 In the previous subsection, we present how to enhance the quality
of the observations. Without any information of multipath delay, the
$L_{C}$ signals in (\ref{eq: y_avg}), $\{y_{n_{w},n_{f}}'[l_{c}],\,l_{c}=0,\cdots,L_{C}-1\}$,
with respect to a certain beam pair $(\tilde{\mathbf{f}}_{n_{f}},\tilde{\mathbf{w}}_{n_{w}})$
are regarded as useful observations. Nevertheless, only $P$ sparse
observations corresponding to the $P$ CIRs are exactly useful. Fortunately,
we can borrow the idea of the analog beam selection in (\ref{eq: ABF})
to find the multipath delay indices because the signals $\{y_{n_{w},n_{f}}'[l_{c}],\,\forall n_{w},n_{f},l_{c}\}$
are represented in the discrete delay-angle domain, where $l_{c}$
and $(n_{w},n_{f})$ respectively denote the delay- and angle-domain
indices. Accordingly, the multipath delay estimation can be stated
as the following problem: given $\{y_{n_{w},n_{f}}'[l_{c}],\,\forall n_{w},n_{f},l_{c}\}$,
one can calculate the sum of the power of $N_{W}N_{F}$ observations
across all steering angles as
\begin{equation}
\begin{alignedat}{1}f[l_{c}] & =\sum_{n_{w}=1}^{N_{W}}\sum_{n_{f}=1}^{N_{F}}\left|y_{n_{w},n_{f}}'[l_{c}]\right|^{2}\end{alignedat}
,
\end{equation}
and solve the constrained maximization problem 
\begin{equation}
\begin{gathered}\hat{l}_{\hat{p}}={\displaystyle \underset{\tiny l_{c}\in\{0,\cdots,L_{C}-1\}\backslash\mathcal{L}}{\arg\text{ }\max}}f[l_{c}],\\
\text{s.t. }\begin{cases}
f[l_{c}]\geq\mu,\\
\mathcal{L}=\{\hat{l}_{n},n=1,\cdots,\hat{p}-1\},
\end{cases}
\end{gathered}
\label{eq: l_hat}
\end{equation}
where $\hat{p}=1,\cdots,\hat{P}$ denotes the path index whose received
power across all steering angles is greater than or equal to a pre-defined
threshold $\mu$, and $\mathcal{L}$ is the set containing the selected
path indices from iteration $1$ to $\hat{p}-1$. Here we consider
the sum of the power of $N_{W}N_{F}$ observations in the angle domain;
therefore the threshold can be simply assumed to be $\mu=N_{W}N_{F}\left(\frac{\sigma_{\xi}^{2}}{M}\right)$.
Since mmWave channels are sparse in nature, the multipath delay indices
can be estimated by using this characteristic to further improve the
performance \cite{Gui2010}. Due to the page limit, we do not provide
more discussion.

According to the estimated delay indices, only $\hat{P}$ refined
observations are used for the analog beam selection problem (assume
$P\leq\hat{P}<L_{C}$ but $\{\hat{l}_{\hat{p}}\;\forall\hat{p}\}$
does not necessarily include $\{l_{p}\:\forall p\}$) and the corresponding
objective function is given by
\begin{equation}
\begin{alignedat}{1}g_{n_{w},n_{f}}'' & =\sum_{\hat{p}=1}^{\hat{P}}\left|y_{n_{w},n_{f}}'[\hat{l}_{\hat{p}}]\right|^{2}\end{alignedat}
.\label{eq: g_exp}
\end{equation}
Similarly, conditioned on the same channel realizations, $\underline{\mathbf{H}}[l]$,
$l=0,\cdots,L-1$, we have the MAE between $g_{n_{w},n_{f}}''$ and
its true value $g_{n_{w},n_{f}}^{N\!F}$ upper bounded by
\begin{equation}
\begin{alignedat}{1}\text{MAE}(\underline{g}_{n_{w},n_{f}}'') & \triangleq\text{E}\left[\left|\underline{g}_{n_{w},n_{f}}''-\underline{g}_{n_{w},n_{f}}^{N\!F}\right|\right]\\
 & \leq\underline{g}_{n_{w},n_{f}}^{N\!F}-\underline{g}_{n_{w},n_{f}}''^{N\!F}+\text{E}\left[\left|\varepsilon_{n_{w},n_{f}}''\right|\right]+\text{E}\left[\nu''\right],
\end{alignedat}
\label{eq: MAE_imp}
\end{equation}
where
\begin{equation}
\underline{g}_{n_{w},n_{f}}''^{N\!F}=\sum_{\hat{p}=1}^{\hat{P}}\left|\underline{y}_{n_{w},n_{f}}^{N\!F}[\hat{l}_{\hat{p}}]\right|^{2}
\end{equation}
and $\underline{g}_{n_{w},n_{f}}''^{N\!F}\leq\underline{g}_{n_{w},n_{f}}^{N\!F}$
with equality iff $\{\hat{l}_{\hat{p}}\:\forall\hat{p}\}\supseteq\{l_{p}\:\forall p\}$.
Furthermore, $\varepsilon_{n_{w},n_{f}}''$ and $\nu''$ are given
as follows:

\begin{equation}
\varepsilon_{n_{w},n_{f}}''\sim\mathcal{N}\left(0,2\left(\frac{\sigma_{\xi}^{2}}{M}\right)\underline{g}_{n_{w},n_{f}}''^{N\!F}\right)\label{eq: epsilon_double_prime}
\end{equation}
and
\begin{equation}
\nu''\sim\Gamma\left(\hat{P},\frac{\sigma_{\xi}^{2}}{M}\right).\label{eq: nu_double_prime}
\end{equation}
Compared with $\varepsilon_{n_{w},n_{f}}'$ and $\nu'$, although
the variance of $\varepsilon_{n_{w},n_{f}}''$ and the shape parameter
of $\nu''$ become smaller, $\text{MAE}(\underline{g}_{n_{w},n_{f}}'')$
is not necessarily less than $\text{MAE}(\underline{g}_{n_{w},n_{f}}')$
if the difference between $\underline{g}_{n_{w},n_{f}}^{N\!F}$ and
$\underline{g}_{n_{w},n_{f}}''^{N\!F}$ is too large. It depends on
the performance of multipath delay estimation.

{\small{}}

\section{Numerical Results\label{sec:Numerical-Results}}

The system parameters used in the simulations are listed below:\vspace{0.15cm}\\
\vspace{0.15cm}%
\begin{tabular}{ll}
Number of antennas & $N_{T}=N_{R}=32$\tabularnewline
Number of RF chains & $N_{RF}=2$\tabularnewline
Number of samples per OFDM symbol & $L=2048$\tabularnewline
CP length & $L_{C}=128$\tabularnewline
Codebook size & $N_{F}=N_{W}=32$\tabularnewline
Number of paths & $P=10$\tabularnewline
\end{tabular}\\
In addition, the effective noise variance is given by $\sigma_{\xi}^{2}=\rho\cdot10^{-\gamma/10}$,
where $\gamma$ (dB) is the SNR. In the codebooks, $32$ steering
angle candidates are: $\left\{ \frac{180^{\circ}}{\pi}\cdot\sin^{-1}\left(\frac{\left(n_{f}-16\right)}{16}\right),\,n_{f}=1,\cdots,32\right\} $
\cite{Chiang2016_WOWMOM}.
\begin{figure}[t]
\centering{}\includegraphics[scale=0.6]{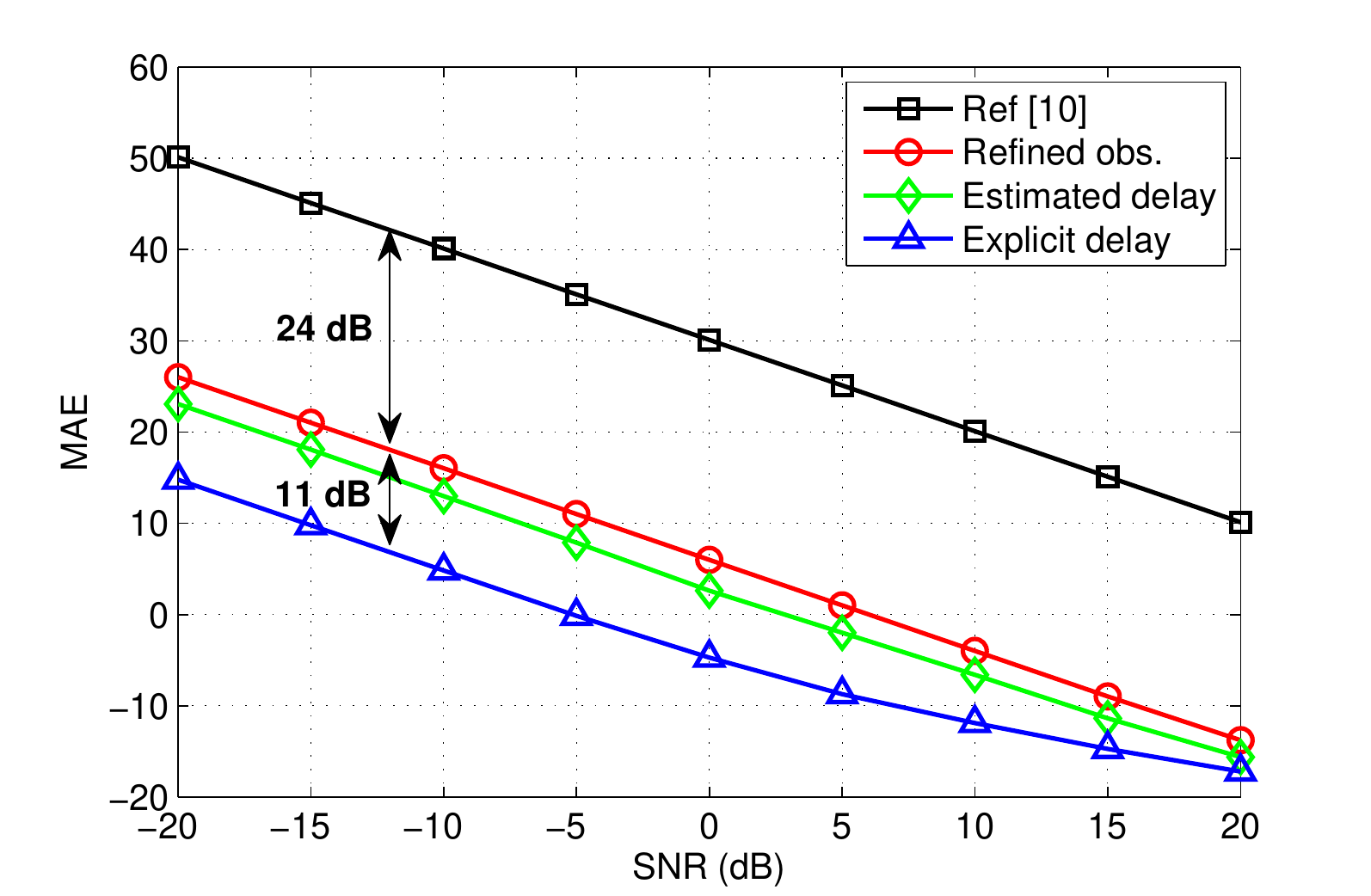}\caption{MAE between energy estimates and their true value, where \textit{Ref}
uses $L=2048$ unrefined observations, and others use $L_{C}=128$
refined observations with and without knowledge of multipath delay.\label{fig: MAE}}
\end{figure}

As discussed in Section \ref{subsec:Problem-statement}, the true
value of the energy yields the optimal solution of the problem in
(\ref{eq: ABF}). Let us denote the indices of the optimal beam pairs
as $(\mathring{n}_{w,n_{r\!f}},\mathring{n}_{f,n_{r\!f}})\:\forall n_{r\!f}$,
and then use the MAE as a performance metric to evaluate the performance
of the proposed and reference methods with respect to the beam pairs
$(\mathring{n}_{w,n_{r\!f}},\mathring{n}_{f,n_{r\!f}})\:\forall n_{r\!f}$.
In Fig. \ref{fig: MAE}, the curves labeled as \textit{Ref}, \textit{Refined
obs}., and \textit{Estimated delay} are respectively calculated by
the following equations:

\vspace{-0.4cm}\begin{subequations}{\small{}}{\small \par}

\begin{flalign}
\text{Ref} & =\frac{1}{N_{RF}}\sum_{n_{r\!f}=1}^{N_{RF}}\text{MAE}(g_{\mathring{n}_{w,n_{r\!f}},\mathring{n}_{f,n_{r\!f}}}),\label{eq: MAE1}\\
\text{Refined obs.} & =\frac{1}{N_{RF}}\sum_{n_{r\!f}=1}^{N_{RF}}\text{MAE}(g_{\mathring{n}_{w,n_{r\!f}},\mathring{n}_{f,n_{r\!f}}}'),\label{eq: MAE2}\\
\text{Estimated delay} & =\frac{1}{N_{RF}}\sum_{n_{r\!f}=1}^{N_{RF}}\text{MAE}(g_{\mathring{n}_{w,n_{r\!f}},\mathring{n}_{f,n_{r\!f}}}''),\label{eq: MAE3}
\end{flalign}
\end{subequations}where the energy estimate in (\ref{eq: MAE1})
is equivalent to the sum of the power of observations across all subcarriers,
which is the objective function of the frequency-domain analog beam
selection problem in \cite{Chiang2018_ICC_FD}. 

From (\ref{eq:  MAE}) and (\ref{eq: MAE_avg}), we can find the upper
bounds of (\ref{eq: MAE1}) and (\ref{eq: MAE2}), and they are dominated
by the gamma distributed random variables when the values of shape
and scale parameters are large. As a result, (\ref{eq: MAE1}) and
(\ref{eq: MAE2}) can be approximated by\begin{subequations}
\begin{alignat}{1}
(\ref{eq: MAE1}) & \approx\text{E}\left[\nu\right],\\
(\ref{eq: MAE2}) & \approx\text{E}\left[\nu'\right].
\end{alignat}
\end{subequations}and therefore the difference in MAE between \textit{Ref}
and \textit{Refined obs.} is given by 
\[
10\log_{10}\left(\frac{\text{E}\left[\nu\right]}{\text{E}\left[\nu'\right]}\right)=10\log_{10}\left(M^{2}\right)=24.08\,\text{dB}.
\]

\begin{figure}[t]
\centering{}\includegraphics[scale=0.6]{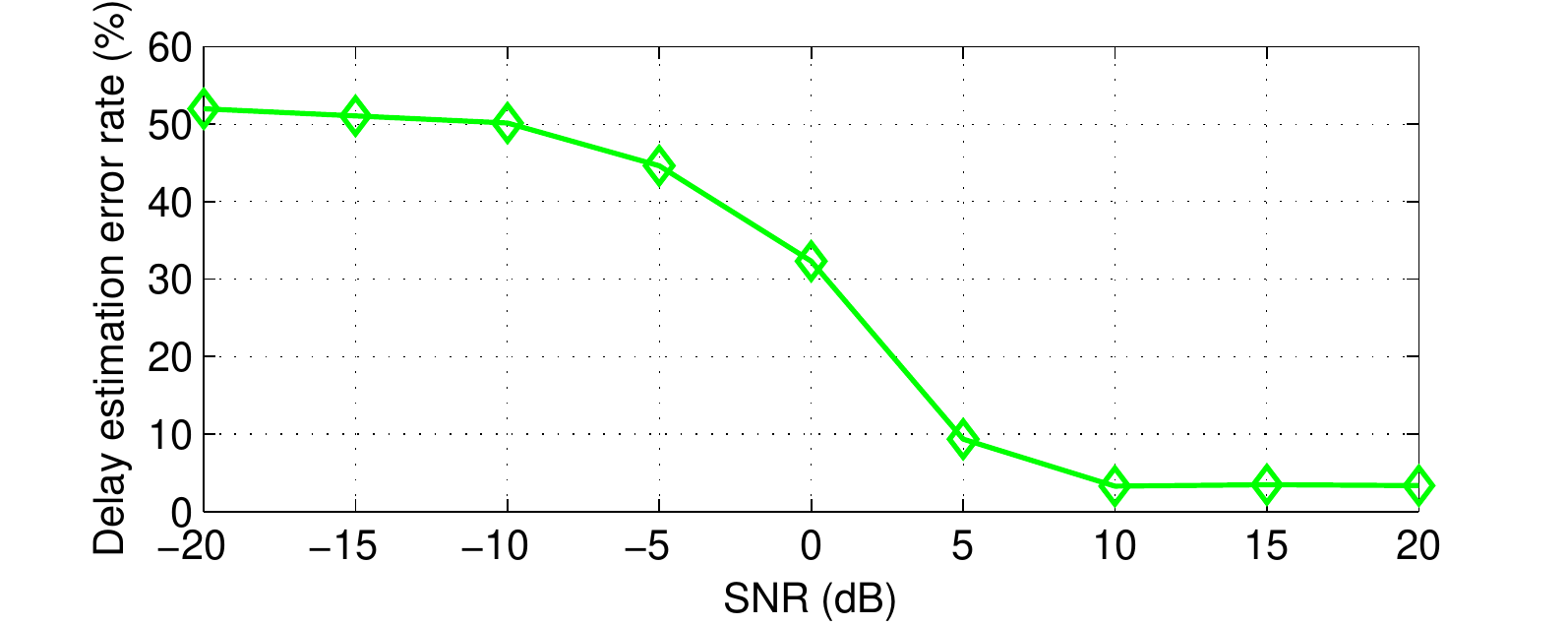}\caption{Estimation error rate of $P$ delay indices in  curve \textit{Estimated
delay} in Fig. \ref{fig: MAE}.\label{fig: delay_error}}
\end{figure}

In (\ref{eq: MAE3}), if we only use $\hat{P}$ refined observations
associated with $\hat{P}$ estimated delay indices (the estimation
error rate is shown in Fig. \ref{fig: delay_error}), the MAE can
be reduced by $3$ dB compared with \textit{Refined obs}., see  curve
\textit{Estimated delay}. Ideally, if the set containing $\hat{P}$
estimated delay indices is equal to $\{l_{p}\:\forall p\}$, following
from (\ref{eq: MAE_imp}), the corresponding MAE is upper bounded
by
\begin{equation}
\begin{alignedat}{1} & \text{MAE}(\underline{g}_{n_{w},n_{f}}''|_{\text{given }\{l_{p}\,\forall p\}})\\
 & \leq\text{E}\left[\left|\varepsilon_{n_{w},n_{f}}''|_{\text{given }\{l_{p}\,\forall p\}}\right|\right]+\text{E}\left[\nu''|_{\text{given }\{l_{p}\,\forall p\}}\right]\\
 & =\text{E}\left[\left|\varepsilon_{n_{w},n_{f}}'\right|\right]+\text{E}\left[\nu''|_{\text{given }\{l_{p}\,\forall p\}}\right]
\end{alignedat}
\label{eq: MAE_exp}
\end{equation}
where
\begin{equation}
\nu''|_{\text{given }\{l_{p}\,\forall p\}}\sim\Gamma\left(P,\frac{\sigma_{\xi}^{2}}{M}\right),\label{eq: nu_double_prime-2}
\end{equation}
and the simulation results are shown in  curve \textit{Explicit delay}
calculated by
\begin{equation}
\text{Explicit delay}=\frac{1}{N_{RF}}\sum_{n_{r\!f}=1}^{N_{RF}}\text{MAE}(\underline{g}_{\mathring{n}_{w,n_{r\!f}},\mathring{n}_{f,n_{r\!f}}}''|_{\text{given }\{l_{p}\,\forall p\}}).\label{eq: MAE4}
\end{equation}
In the low SNR regime, $\text{MAE}(\underline{g}_{n_{w},n_{f}}''|_{\text{given }\{l_{p}\,\forall p\}})$
is dominated by the gamma distributed random variable as well. Hence,
the difference in MAE between \textit{Refined obs.} and \textit{Explicit
delay} approximates to

{\small{}\vspace*{-0.2cm}
\begin{equation}
10\log_{10}\left(\frac{\text{E}\left[\nu'\right]}{\text{E}\left[\nu''|_{\text{given }\{l_{p}\,\forall p\}}\right]}\right)=10\log_{10}\left(\frac{L}{MP}\right)=11.07\,\text{dB}.\label{eq: diff_refine_obs_and_explicit}
\end{equation}
}When the SNR increases, (\ref{eq: MAE4}) is not only dominated by
the gamma distributed random variable, so the difference between \textit{Refined
obs. }and\textit{ Explicit delay} cannot be simply approximated by
(\ref{eq: diff_refine_obs_and_explicit}).

In Fig. \ref{fig: delay_error}, it shows the estimation error rate
of $P$ delay indices in  curve \textit{Estimated delay} in Fig. \ref{fig: MAE}.
As mentioned in Section \ref{subsec:Further-refine-observations},
since the exact number $P=10$ of paths is not available, we try to
find $\hat{P}$ paths whose sum of the received power across all steering
angles are greater than or equal to the pre-defined threshold $\mu=\left(\frac{\sigma_{\xi}^{2}}{M}\right)N_{W}N_{F}$.
When $\text{SNR}<0\text{ dB}$, the delay estimation error rate of
more than $30\%$ leads to an MAE reduction of approximately $3$
dB, compared with \textit{Refined obs}. On the other hand, when $\text{SNR}\geq10\text{ dB}$,
the delay estimation error rate approximates to zero. However, the
gap between \textit{Estimated delay }and \textit{Explicit delay} in
Fig. \ref{fig: MAE} is still quite obvious, which means that $\hat{P}\gg P$
and therefore not only the useful observations but also a large number
of noise signals are reserved. The delay estimation approach can be
further enhanced by, for example, modifying the threshold; nevertheless,
it is beyond the scope of this paper.

{\small{}}
\begin{figure*}[t]
{\small{}
\begin{equation}
\begin{alignedat}{1}\underline{g}_{n_{w},n_{f}} & =\sum_{l=0}^{L-1}\left|\sqrt{\rho}\,\tilde{\mathbf{w}}_{n_{w}}^{H}\underline{\mathbf{H}}[l]\tilde{\mathbf{f}}_{n_{f}}+\xi_{n_{w},n_{f}}[l]\right|^{2}\\
 & =\sum_{l=0}^{L-1}\left|\underline{y}_{n_{w},n_{f}}^{N\!F}[l]+\xi_{n_{w},n_{f}}[l]\right|^{2}\\
 & =\sum_{l=0}^{L-1}\mathfrak{R}\left(\underline{y}_{n_{w},n_{f}}^{N\!F}[l]+\xi_{n_{w},n_{f}}[l]\right)^{2}+\mathfrak{I}\left(\underline{y}_{n_{w},n_{f}}^{N\!F}[l]+\xi_{n_{w},n_{f}}[l]\right)^{2}\\
 & =\underset{\triangleq\underline{g}_{n_{w},n_{f}}^{N\!F}}{\underbrace{\sum_{l=0}^{L-1}\mathfrak{R}\left(\underline{y}_{n_{w},n_{f}}^{N\!F}[l]\right)^{2}+\mathfrak{I}\left(\underline{y}_{n_{w},n_{f}}^{N\!F}[l]\right)^{2}}}+\underset{\triangleq\varepsilon_{n_{w},n_{f}}}{\underbrace{\sum_{l=0}^{L-1}2\mathfrak{R}\left(\underline{y}_{n_{w},n_{f}}^{N\!F}[l]\right)\mathfrak{R}\left(\xi_{n_{w},n_{f}}[l]\right)+2\mathfrak{I}\left(\underline{y}_{n_{w},n_{f}}^{N\!F}[l]\right)\mathfrak{I}\left(\xi_{n_{w},n_{f}}[l]\right)}}\\
 & \quad\,+\underset{\triangleq\nu}{\underbrace{\sum_{l=0}^{L-1}\mathfrak{R}\left(\xi_{n_{w},n_{f}}[l]\right)^{2}+\mathfrak{I}\left(\xi_{n_{w},n_{f}}[l]\right)^{2}}}\\
 & =\underline{g}_{n_{w},n_{f}}^{N\!F}+\varepsilon_{n_{w},n_{f}}+\nu,
\end{alignedat}
\label{eq: g_underline}
\end{equation}
}{\small \par}

{\footnotesize{}\hrulefill}{\footnotesize \par}
\end{figure*}
{\small \par}

\section{Conclusion}

The mmWave channel sparsity in the delay domain is widely acknowledged
as a powerful cue for analog beam selection. Different to the conventional
methods addressing the feature in the frequency domain, this paper
presents a new perspective in the delay domain and shows that the
significant observations used for the analog beam selection are also
sparse. To improve the quality of the observations, we propose a solution
that transmits the periodic training sequence of length equal to a
CP length. An arithmetic mean can accordingly reduce the noise variance
to refine the observations. Then based on the refined signals represented
in the delay-angle domain, the sparse significant observations can
be simply captured by finding the maximum term in the sum of the power
of the refined signals across angle.

\section{Appendix}

\subsection{Proof of Theorem \ref{thm: MAE}}

Given channel matrices $\underline{\mathbf{H}}[l]$, $l=0,\cdots,L-1$,
the objective function $g_{n_{w},n_{f}}$ in (\ref{eq: g_ori}) becomes
(\ref{eq: g_underline}), where the first term $\underline{g}_{n_{w},n_{f}}^{N\!F}$
is a constant, the second term $\varepsilon_{n_{w},n_{f}}$ has a
normal distribution with mean zero and variance $2\sigma_{\xi}^{2}\underline{g}_{n_{w},n_{f}}^{N\!F}$,
$\varepsilon_{n_{w},n_{f}}\sim\mathcal{N}(0,2\sigma_{\xi}^{2}\underline{g}_{n_{w},n_{f}}^{N\!F})$,
and the third term $\nu$ has a gamma distribution with the shape
parameter $L$ and scale parameter $\sigma_{\xi}^{2}$, $\nu\sim\Gamma(L,\sigma_{\xi}^{2})$.
Therefore, the MAE between $\underline{g}_{n_{w},n_{f}}$ and $\underline{g}_{n_{w},n_{f}}^{N\!F}$
(denoted as $\text{MAE}(\underline{g}_{n_{w},n_{f}})$) is given by
\[
\begin{alignedat}{1}\text{MAE}(\underline{g}_{n_{w},n_{f}}) & \triangleq\text{E}\left[\left|\underline{g}_{n_{w},n_{f}}-\underline{g}_{n_{w},n_{f}}^{N\!F}\right|\right]\\
 & =\text{E}\left[\left|\underline{g}_{n_{w},n_{f}}^{N\!F}+\varepsilon_{n_{w},n_{f}}+\nu-\underline{g}_{n_{w},n_{f}}^{N\!F}\right|\right]\\
 & =\text{E}\left[\left|\varepsilon_{n_{w},n_{f}}+\nu\right|\right]\\
 & \leq\text{E}\left[\left|\varepsilon_{n_{w},n_{f}}\right|\right]+\text{E}\left[\nu\right].
\end{alignedat}
\]
\\

\section*{Acknowledgment}

The research leading to these results has received funding from the
European Union\textquoteright s Horizon 2020 research and innovation
programme under grant agreement No. 671551 (5G-XHaul) and the TUD-NEC
project \textquotedblleft mmWave Antenna Array Concept Study\textquotedblright ,
a cooperation project between Technische Universität Dresden (TUD),
Germany, and NEC, Japan.

\bibliographystyle{IEEEtran}
\bibliography{reference}

\end{document}